\documentclass[12pt,preprint]{aastex}

\slugcomment{Submitted: Astrophysical Journal, September 2008}

\shorttitle{Mid-IR PL Relations}
\shortauthors{Madore {\it et al.}}

\begin{document}


\title{The Cepheid Period-Luminosity Relation at Mid-Infrared
Wavelengths: II. Second-Epoch LMC Data}


\author{\bf Barry F. Madore, Wendy L. Freedman, Jane Rigby \\ S. E. Persson,  Laura Sturch \& Violet Mager}
\affil{Observatories of the  Carnegie Institution of Washington \\ 813
Santa
Barbara St., Pasadena, CA ~~91101}
\email{wendy@ociw.edu, barry@ociw.edu, jrigby@ociw.edu,  
persson@ociw.edu, lsturch@ociw.edu, vmager@ociw.edu}

%% are identified by the \altaffilmark after each name.  Specify  alternate


\begin{abstract}
We present revised and improved mid-infrared Period-Luminosity (PL)
relations for Large Magellanic Cloud (LMC) Cepheids based on
double-epoch data of 70 Cepheids observed by {\it Spitzer} at 3.6,
4.5, 5.8 and 8.0~$\mu$m.  The observed scatter at all wavelengths is
found to decrease from $\pm$0.17 mag to $\pm$0.14 mag, which is fully
consistent with the prediction that the total scatter is made up of
roughly equal contributions from random sampling of the light curve
and nearly-uniform samplings of stars across the instability strip.
It is calculated that the Cepheids in this sample have a full
amplitude of about 0.4~mag and that their fully-sampled, time-averaged
magnitudes should eventually reveal mid-infrared PL relations that
each have intrinsic scatter at most at the $\pm$0.12~mag level, and as
low as $\pm$0.08~mag after correcting for the tilt of the LMC.

\end{abstract}

\vfill\eject
\section{INTRODUCTION}

In an earlier paper (Freedman et al. 2008, hereafter Paper I) we
presented a preliminary calibration of the mid-infrared LMC Cepheid
Period-Luminosity (PL) relations at 3.6, 4.5, 5.8 and 8.0~$\mu$m
followed by an analysis of the wavelength dependence of the slopes and
scatter in those relations. Those observations were derived from data
serendipitously obtained by {\it Spitzer} during the course of the
{\it Surveying the Agents of a Galaxy's Evolution} (SAGE) Project
(Meixner {\it et al.} 2006a). The main conclusion was that even
single-phase data (for known Cepheids with published periods) provided
PL relations that had a mean scatter of only about $\pm$0.17~mag,
largely independent of wavelength. A plausible case was made that about
half of the scatter was due to the random-phase nature of single
observations (sampling light curves with a full amplitude of
0.3-0.4~mag), and that the other half of the scatter was being
contributed by the intrinsic width of the instability strip. The
latter was then self-consistently estimated to have an edge-to-edge
width of 0.3-0.4~mag.  Here we consider the effects of doubling the
number of observations of the same (fixed size) sample of 70
long-period LMC Cepheids as were examined in Paper I.

The second (and last) installment of the combined and revised
catalog\footnote{Available from IRSA at
http://irsa.ipac.caltech.edu/data/SPITZER/SAGE/} of point sources in
the LMC has been released by the SAGE team. We have taken 4-band
photometry from this catalog for epochs 1 and 2; the epoch 1
photometry was replaced by the SAGE team for consistency.  This
provides us with 560 individual observations of 70 Cepheids spread
over 4 wavelengths and separated in time by about 7 months. The quoted
precision of the individual data points varies systematically with
period/magnitude, ranging from $\pm$0.01~mag for the brightest
(longest-period) Cepheids at about 80 days, up to $\pm$0.10 mag for
the faintest (short-period) Cepheids in our sample at about 6 days.
The individual observations for both epochs are listed in Table 1,
revising and updating our earlier tabulation in Freedman et
al. (2008).

Our sample is based on the Persson et al. (2004) subset of
well-observed LMC Cepheids selected to be uncrowded. Significantly
larger samples (200 to nearly 600 stars, depending on wavelength) have
been studied by Ngeow \& Kanbur (2008). Their analysis results in some very
different conclusions from ours. We comment briefly on these differences and
explore their probable origin at the end of Sections 2 and 3.

\section{IRAC Mid-Infrared Period-Luminosity Relations}

The four mid-IR period-luminosity relations, presenting all 560
random-phase observations of the 70 Cepheids, are given in Figures 1
and 2.  Details of the SAGE observations are given in Paper I; here we
simply note that for some of the Cepheids the data for the reprocessed
photometry differs slightly from the previous release used in Paper
I. In this paper we use the most recently processed photometry for
both epochs.  The solid line gives a least-squares fit to the plotted
data. The dashed lines mark the two-sigma limits on the scatter in
each of the PL relations taken from Paper I. Despite the factor-of-two
increase in the number of observations it is interesting to note that
virtually all of the new data points fall within the previous
limits. This was to be expected if the observed range of the
distribution is being defined by random sampling of finite-amplitude
light curves. If the scatter were instead dominated by random errors
in the photometry alone, the range would increase with sample size; it
does not. We can then with some confidence go on to interpret the
scatter physically.

Averaging the two-epoch photometry for individual Cepheids 
significantly decreases the scatter about the mean PL relation. This
is because averaging brings us closer to the time-averaged mean
magnitude of the individual Cepheids (by damping out the phase-induced
excursions.)

How much of the total scatter is due to the random sampling of the
light curves, and how much is due to the intrinsic width of the
Cepheid instability strip, projected into the period-luminosity plane?
In Paper I we suggested that the split was 50:50. We can now
quantitatively test that prediction.

\subsection{Period-Luminosity Fits}

Weighted, least-squares, linear fits to each of the four mid-IR data
sets are given in Section 3.  The slopes, zero points, respective errors
and $rms$ scatter, listed for each bandpass, are consistent with our
previous results from Paper I. The slopes at each wavelength all have
values of around -3.4, with a slight trend of increasingly negative
values towards longer wavelengths. The scatter around each of the fits is
relatively constant at $\pm$0.16-0.17~mag; and from filter to filter
the scatter of individual data points is highly correlated (see Figures
3-5 and discussion below).

\subsection{Correlations in the Scatter}

In Figure 5 we show the highly correlated nature of the magnitude
differences in a given band plotted against the magnitude differences
for the same stars, but at the other wavelengths. Clearly the observed
differences are not random noise, but rather due to the physically
correlated changes generated by light variation of the Cepheids
themselves. Since there appears to be no significant deviation from a
one-to-one correspondence in these variations, this is consistent with
the dominant contributor in each of these bands being due to radius
variations, as expected at these long wavelengths (Madore \& Freedman
1991).

If the magnitude differences in the two observations are purely a
result of changes intrinsic to the Cepheid during its cycle then the
distribution of those differences should be consistent with our
expected mid-infrared light curve and that distribution should reach a
maximum at the full amplitude. Consulting the lower right panel in
Figure 8 of  Madore \& Freedman (2005) which shows the marginalized
distribution of amplitudes for a two-epoch, random sampling of a
Cepheid light curves, shows that the expected distribution function
should be triangular in form with its peak at zero difference between
consecutive observations and its minimum at a value corresponding to
(the unlikely, two-point, chance-sampling of) the full amplitude of
the lightcurve. Examination of Figure 6 shows that the observed
distributions of differences (upper histograms) correspond to this
prediction and that the most probable amplitude for these Cepheids at
all wavelengths is 0.4-0.5 mag. The lower portion of the main plot
showing each of the individual differences (with symbols encoded by
wavelength) indicates that there may be a slow decline of mean/maximum
amplitude with decreasing period from a maximum amplitude of 0.5 mag
around 40 days, falling to a maximum amplitude of 0.3 mag below 10
days.

We are now in a position to test the assumption made in Paper I that
the measured variance in the single-phase PL relation is shared
equally between the random sampling of the instability strip and the
phase sampling of the individual light curve where the width of the
instability strip (0.40~mag) contributes a scatter of $\pm$0.12~mag
and the amplitudes of the individual Cepheids (also 0.4~mag
peak-to-peak) contribute another $\pm$0.12~mag of scatter.

That is, if $\sigma_1$ is the scatter around the PL relation measured
for the single-epoch data, and $\sigma_2$ is the scatter measured for
the time-averaged two-epoch data then:

 $\sigma_1^2 = \sigma_{strip}^2 + \sigma_{cepheid}^2$ 

\par

 $\sigma_2^2 = \sigma_{strip}^2 + \sigma_{cepheid}^2/2$ 

\noindent
Thus 

$ \sigma_{cepheid}^2 = 2(\sigma_1^2 - \sigma_2^2)$

\noindent
and 

$ \sigma_{strip}^2 = 2\sigma_2^2 - \sigma_1^2$

Converting observed scatter to intrinsic width (i.e., $ \sigma_{strip}
\times \sqrt{N}$) gives strip widths of 0.36, 0.37, 0.36 and 0.37~mag
at 3.6, 4.5, 5.8 and 8.0 $\mu$m, respectively, and average Cepheid
amplitudes of 0.43, 0.45, 0.43 and 0.43~mag in those same respective
bandpasses. These amplitudes correspond well to the
independently-estimated amplitudes derived from the
magnitude-difference histograms discussed above.

What this means in practice is that, given the choice between
observing more Cepheids or observing more phase points for the same
set of Cepheids in order to reduce the error on the distance modulus,
increasing the number of observations per Cepheid is initially more
effective given the larger amplitude of the stars compared to the
intrinsic width of the instability strip. The advantage of numbers of
observations over numbers of stars is about 3:2 in the variance, which
scales inversely with N.

The short-wavelength residuals reported by Ngeow \& Kanbur (2008)
based on a completely different and considerably larger sample of 200
to 600 LMC Cepheids (depending on the wavelength), are significantly
smaller than the values given here.  For instance, they give (their
Tables 2 \& 4) a scatter about the mean of $\pm$0.10~mag for the 3.6
and 4.5$\mu$m PL relations. Following the line of argument given above
such small scatter (if real) would reduce both the amplitudes and the
width of the instability strip to implausibly small values. Our data
do not support those conclusions. It is possible that the procedures
used by Ngeow \& Kambur in an attempt to remove outliers resulted in
this remarkably low dispersion.

%\vfill\eject
\section{Absolute Calibration}

In keeping with Paper I, we adopt a true distance to the
Large Magellanic Cloud of (m-M)$_o$ = 18.50~mag and a mean reddening
to the Cepheids of E(B-V) = 0.10~mag (Freedman {\it et al.}, 2001).
Applying this correction for the true distance modulus and applying
extinction corrections (0.04 to 0.01~mag), we
derive the following updated absolute calibrations for the Cepheid
Period-Luminosity relations at mid-infrared wavelengths:

$$<M>_{3.6} = -3.40 (log(P) - 1.0) ~[\pm 0.02] - 5.81 ~[\pm 0.03] \ \ \ \ \ \sigma_{3.6} = \pm0.135$$

$$<M>_{4.5} = -3.35 (log(P) - 1.0) ~[\pm 0.02] - 5.76 ~[\pm 0.03] \ \ \ \ \ \sigma_{4.5} = \pm0.141$$

$$<M>_{5.8} = -3.44 (log(P) - 1.0) ~[\pm 0.03] - 5.81 ~[\pm 0.04] \ \ \ \ \ \sigma_{5.8} = \pm0.137$$

$$<M>_{8.0} = -3.49 (log(P) - 1.0) ~[\pm 0.03] - 5.81 ~[\pm 0.04] \ \ \ \ \ \sigma_{8.0} = \pm0.139$$

The last entry in each line is the scatter measured about the
preceding regression line.  The above-quoted slopes and their
generally monotonic increase with wavelength (consistent with and
confirming Figure 4, Paper I) are in conflict with the generally lower
values and especially the puzzling reversal of the slopes with
wavelength given by Ngeow \& Kanbur (2008). We speculate that their
increased sample size brings with it crowding and other photometric
errors that affect their solutions. Certainly the
dramatic increase in dispersion around their solutions at longer
wavelengths must be the result of photometric uncertainties and not
due to (or expected from) physical processes intrinsic to the Cepheids
themselves.

However, we note that there is still two additional sources of
identifiable variance in these solutions: (1) back-to-front geometric
effects due to the tilt of the LMC and (2) a spread of metallicity
amongst the LMC Cepheids themselves. The analysis of Persson et
al. (2004), given in their Table 6, clearly indicates that tilt
contributes about $\pm$0.08~mag of scatter to all solutions using this
Cepheid sample. Removing that geometric term would suggest that the
underlying (time-averaged) mid-infrared PL relations have an intrinsic
scatter of $\pm$0.08~mag. The compilation of measured atmospheric
metallicities of individual LMC Cepheids by Romaniello et al. (2008)
shows a spread of 0.5~dex in [Fe/H]. Any impact of this dispersion of
metallicity on the mid-infrared magnitudes of these LMC Cepheids must
be fully contained within the $\pm$0.08~mag of residual
scatter. However this scatter cannot be totally due to metallicity
since there must ultimately be contributions due to mean-radius and
mean-temperature variations of the Cepheids across the instability
strip. Each of these second-order effects and any others that have not
yet been explicitly considered (residual crowding effects and
unresolved binary stars in the sample, for example) all must be
contained in the residual scatter of $\pm$0.08~mag.

\section{Conclusions and Future Prospects}

The total magnitude width of the LMC Cepheid instability strip at
mid-infrared wavelengths is found to be 0.36~mag (uncorrected for
the LMC tilt). This translates into an $rms$ scatter of $\pm$0.11~mag
around any one of the four time-averaged, mid-infrared PL
relations. In order to obtain a distance modulus statistically good to
$\pm$0.01 mag would require a sample of approximately 100 Cepheids. We
further calculate that the average mid-IR amplitude of Cepheids in our
sample is (peak-to-peak) 0.43~mag. To obtain time-averaged mean
magnitudes for individual Cepheids good to $\pm$0.02~mag would then
require on the order of 36 randomly-phased observations.

Testing the mid-infrared PL relations for their sensitivity to
metallicity, and providing an absolute calibration that is independent
of the LMC distance scale are the obvious next steps in this process.
However, Spitzer alone is certainly capable of detecting Cepheids out
to 1-2~Mpc using reasonable (1-2 hour) integration times, but crowding
from AGB and extended-AGB stars is likely to be the real limiting
factor. Applications to galaxies appreciably beyond the Local Group
and their use in shoring up the extragalactic distance scale must
await JWST.

\vfill\eject
\noindent
\centerline{\bf References \rm}
\vskip 0.1cm
\vskip 0.1cm

\par\noindent
Freedman, W.~L., {\it et al.} 2001, \apj, 553, 47

\par\noindent Freedman, W.~L., Madore, B.F., Rigby, J., Persson, S.E.,
\& Sturch, L. 2008, \apj, 679, 71

\par\noindent
Madore, B.~F., \& Freedman, W.~L. 2005, \pasp, 103, 933

\par\noindent
Madore, B.~F., \& Freedman, W.~L. 2005, \apj , 630, 1054

\par\noindent
Ngeow, C., \& Kanbur, S.M. 2008, \apj, 679, 76

\par\noindent
Meixner, M., {\it et al.} 2006, \aj, 132, 2268

\par\noindent 
Persson, S.E., Madore, B.F., Krzeminski, W., Freedman,
W.L., Roth, M.  \& Murphy, D.C., \aj, 128, 2239

\par\noindent
Romaniello et al., 2008, \aap, 488, 731

\par\noindent
Schaltenbrand, R. \& Tammann, G.A., 1970, \aap, 7, 289

\par\noindent

\vskip 0.75cm

\vfill\eject
\centerline{\bf Figure Captions}

\includegraphics [width=12cm, angle=270] {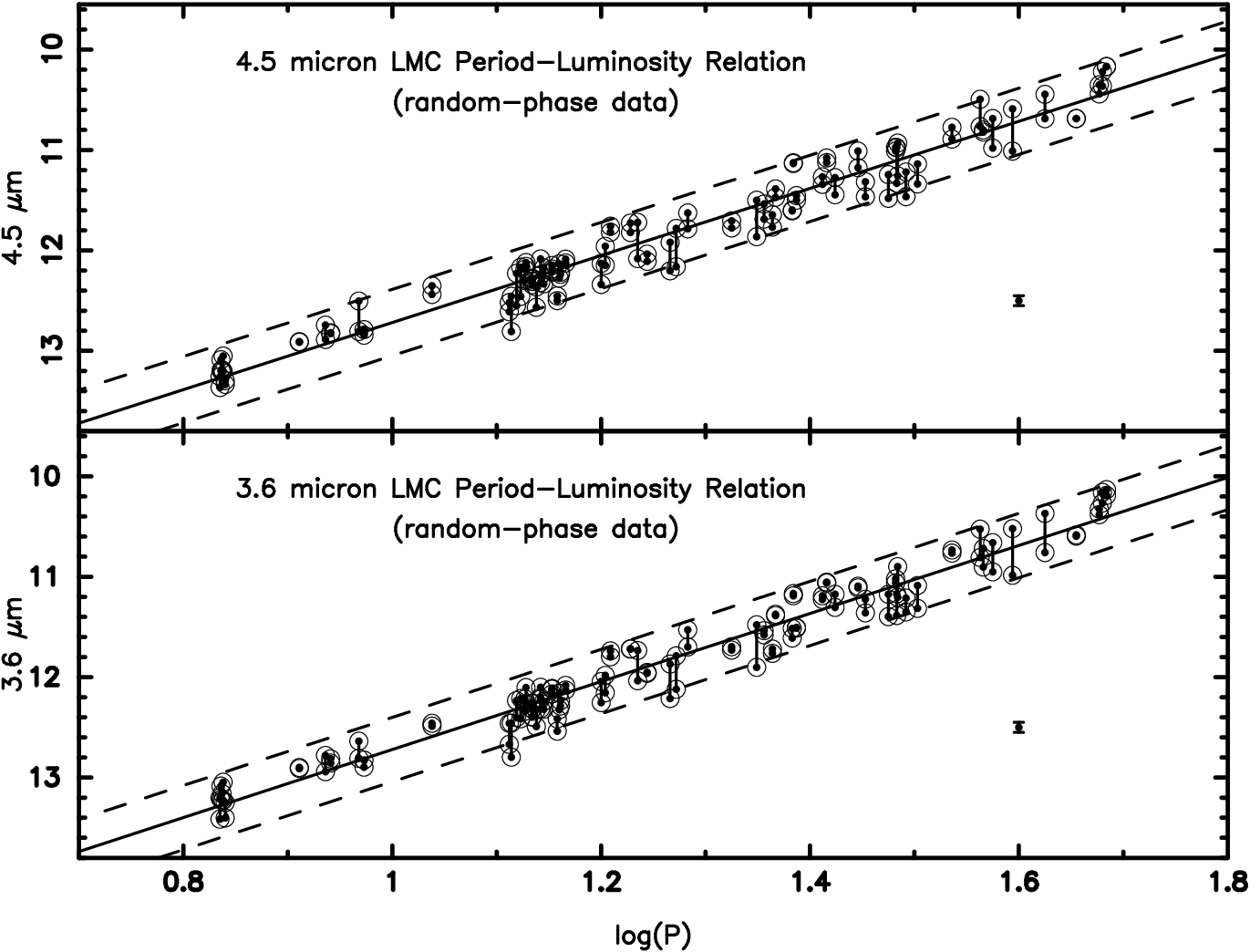}

\par\noindent \bf Fig.~1 \rm -- Random-phase (3.6 and 4.5$\mu$m) IRAC
Period-Luminosity relations for LMC Cepheids plotted over the log(P)
range 0.7 to 1.8.  Observations of the same star seen at different
epochs are joined by solid vertical lines.  The solid line is a
weighted least-squares fit to the data.  The broken lines represent
$\pm$2$\sigma$ (typically $\pm$0.33~mag) bounds on the instability
strip taken from the single-phase data fits, consistent with Paper
I. Note that virtually all of the new data points fall within the old
boundaries: despite a factor of two increase in numbers of data points
the range does not appear to increase.  The sizes of the plotted
symbols are comparable to the typical photometric error on a mid-range
Cepheid, i.e., $\pm$0.05~mag at log(P) = 1.3.  A magnitude error of
$\pm$0.05~mag is shown in the lower right corner of each figure; at
the shortest periods the reported errors on a single observation can
be as high as $\pm$0.10~mag.

\includegraphics [width=12cm, angle=270] {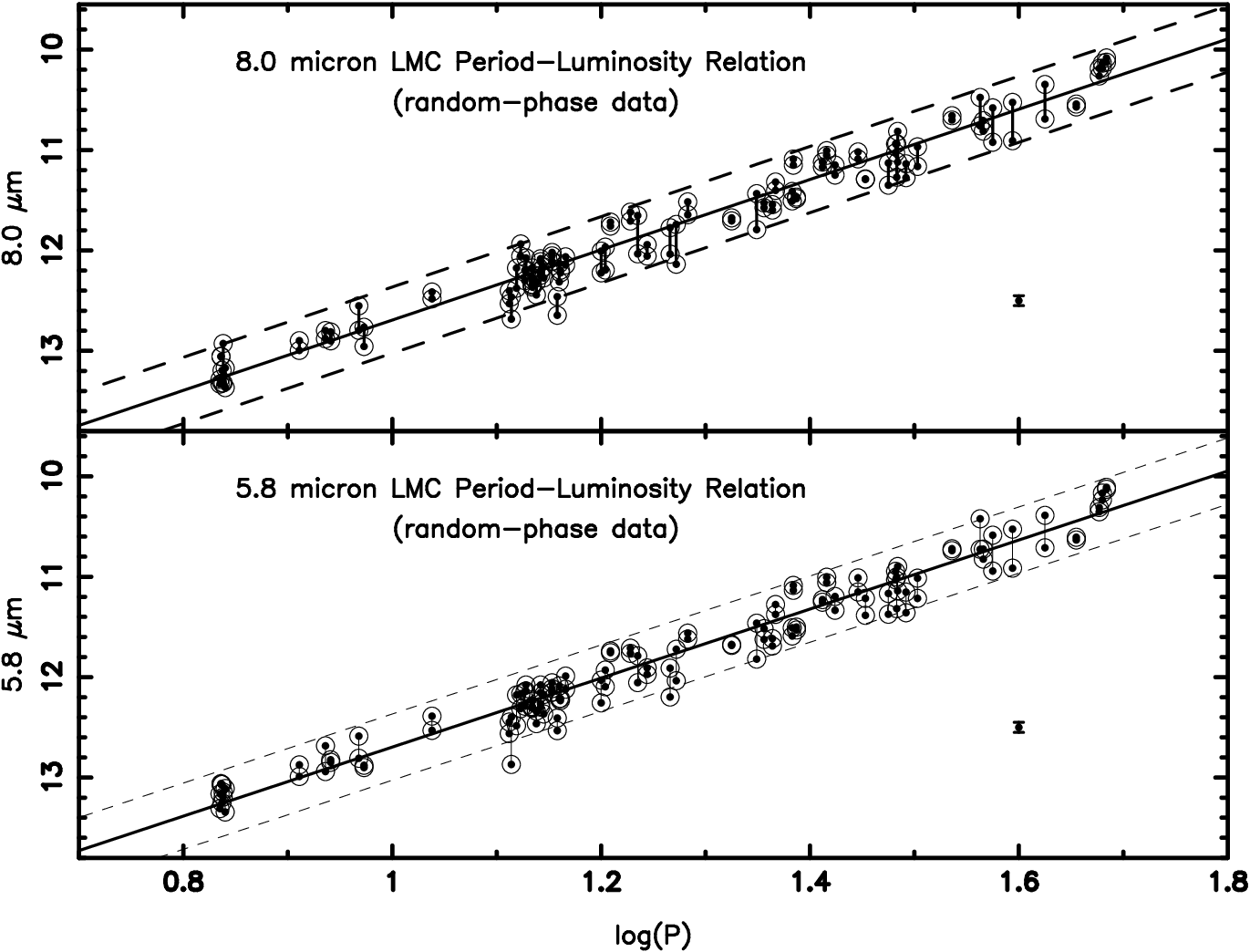} 
\par\noindent \bf Fig.~2 \rm -- The same as Figure 1 except that the
plots are for 5.8 and 8.0$\mu$m IRAC data.

\includegraphics [width=12cm, angle=270] {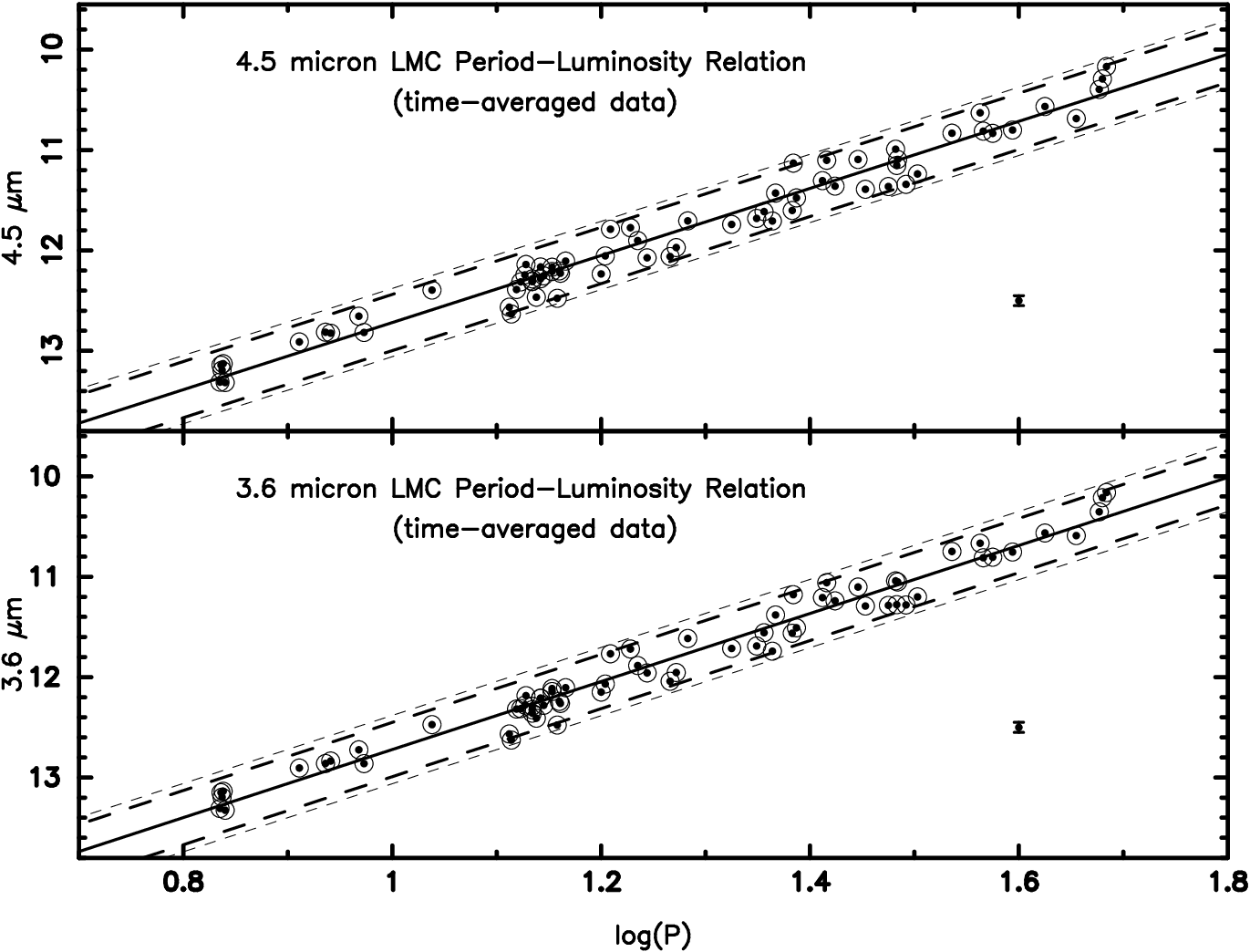} 

\par\noindent \bf Fig.~3 \rm -- Time-Averaged (3.6 and 4.5$\mu$m) IRAC
Period-Luminosity relations for LMC Cepheids plotted over the log(P)
range 0.7 to 1.8. Only two epochs contributed to the average,
nevertheless the scatter drops measurably from around $\pm$0.165~mag
to about $\pm$0.140~mag (see Section 3). The solid line is a weighted
least-squares fit to the data. The thick broken lines are the
$\pm$2$\sigma$ (typically $\pm$0.28~mag) bounds on the instability
strip taken from fits to the plotted time-averaged datasets. The thin
dashed lines are the two-sigma bounds on the instability strip from
Figures 1 \& 2. The shrinkage between the two limits is significant
and is interpreted in the text.  The sizes of the plotted symbols are
comparable to the typical photometric error on a mid-range Cepheid,
i.e., $\pm$0.05~mag at log(P) = 1.3.  A magnitude error of
$\pm$0.05~mag is shown in the lower right corner of each figure; at
the shortest periods the reported errors on a single observation can
be as high as $\pm$0.10~mag.

\includegraphics [width=12cm, angle=270] {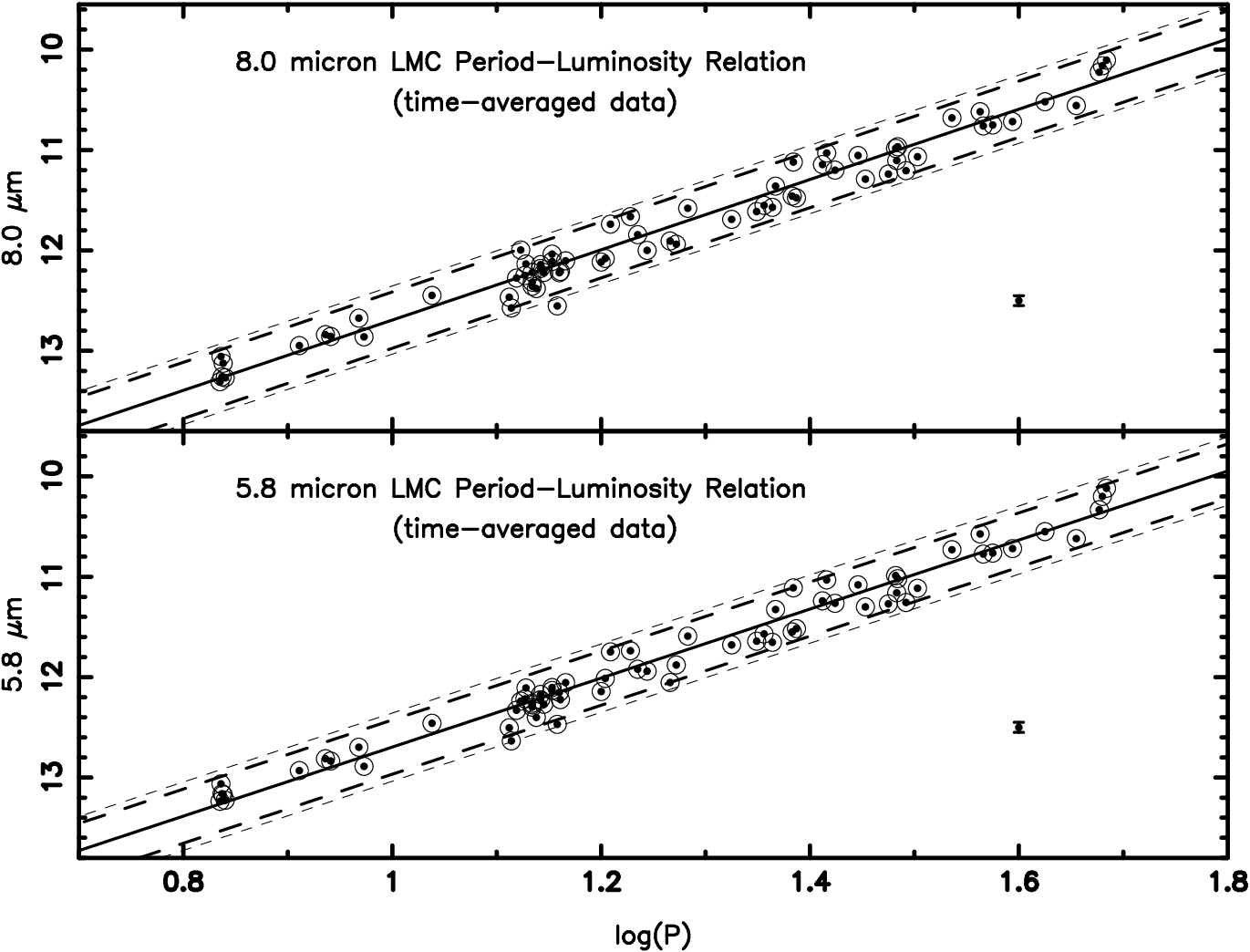}
\par\noindent \bf Fig.~4 \rm -- Time-averaged (5.8 and 8.0$\mu$m) IRAC
Period-Luminosity relations for LMC Cepheids plotted over the log(P)
range 0.7 to 1.8; otherwise see caption to  Figure 3.

\includegraphics [width=14cm, angle=270] {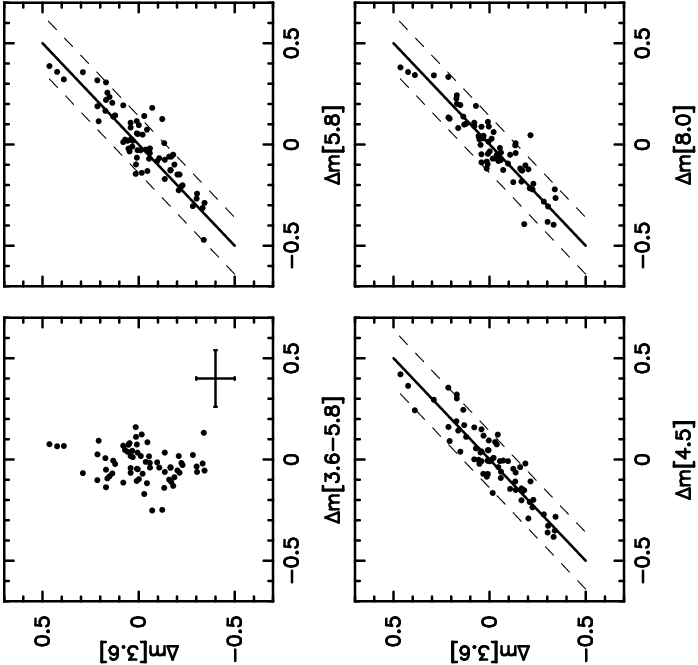} \vfill

\par\noindent \bf Fig. 5 \rm -- Correlated residuals from the PL and
(upper left panel) PC fits. Deviations of individual phase points from
the mean PL regression are shown to be highly correlated across the
respective bandpasses. A unit-slope line is shown for reference, and
scatter around that line is seen to be consistent with the photometric
errors (typically $\pm$0.05~mag) quoted for the individual
observations. Maximum (correlated) excursions have an upper bound of
$\pm$0.4~mag which is probably representative of the full amplitude of
Cepheids at these wavelengths. The upper left panel is illustrative of
the lack of any significant correlation between magnitude and color
residuals, primarily because there is no statistically significant
change (with phase) in the colors of these Cepheids at mid-IR
wavelengths. A typical (two-sigma) error bar is shown in the lower
righthand corner of this plot: the change in magnitude for a given
star is statistically significant, the change in color is not.

\vfill\eject

\includegraphics [width=9cm, angle=270] {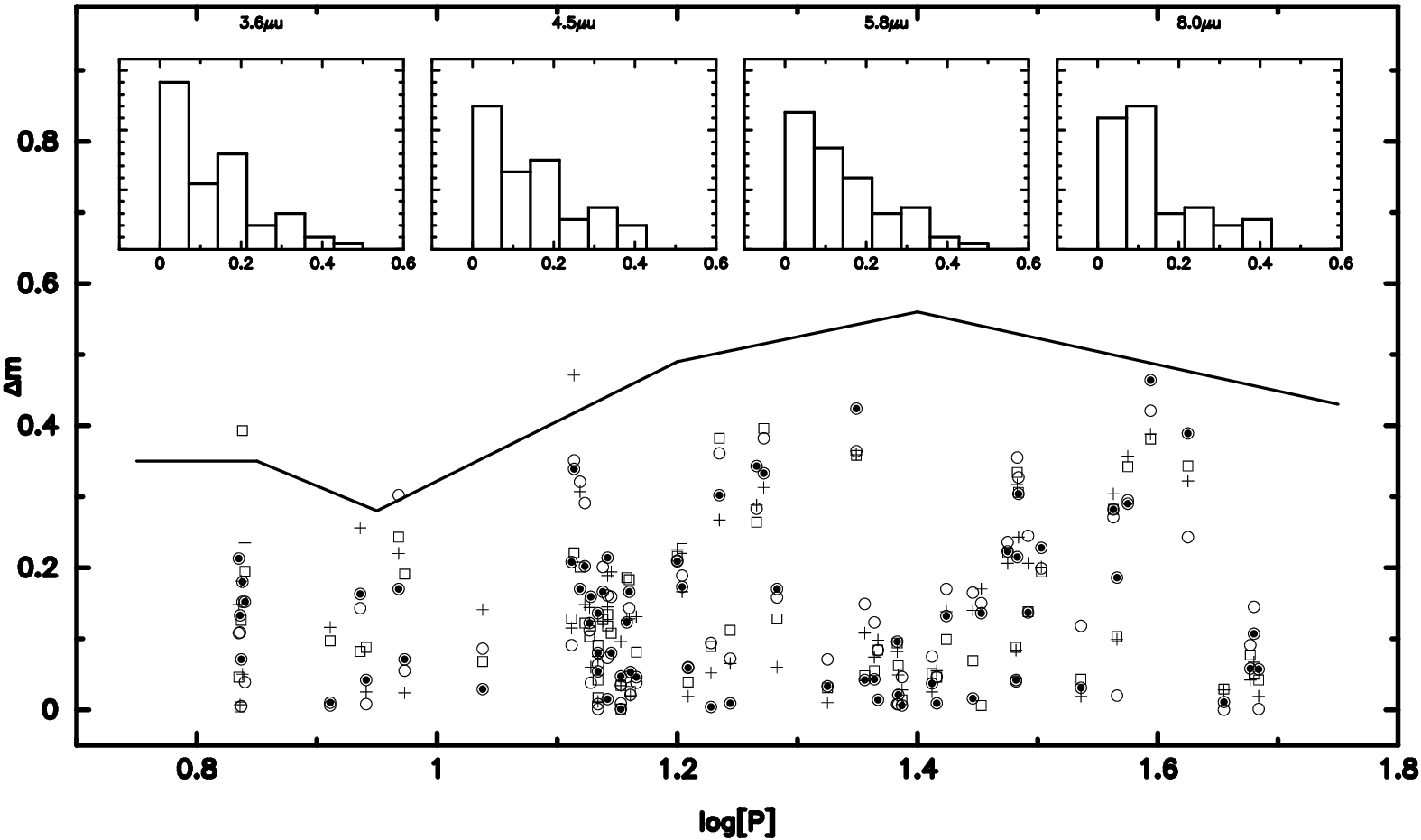} 

\par\noindent \bf Fig.~6 \rm -- The absolute values of magnitude
differences between the two epochs for the same star as a function of
the period. Filled circles are 3.6$\mu$m amplitudes, open circles are
4.5$\mu$m data.  Crosses and squares are magnitude differences for 6.8
and 8.0$\mu$m data, respectively. Note the possible increase in the
maximum difference as a function period, consistent with the scaled
upper envelope to the optical Period-Amplitude relation from
Schaltenbrand \& Tammann (1970), shown the series of broken straight
lines.  The histograms in the upper part of the figure represent the
marginalized and binned amplitude data for each of the four IRAC bands
increasing in wavelength from left to right. As predicted from
simulations (Madore \& Freedman 2005), the form of the histograms
should be trianglar, as is observed; and the x intercept should be the
full amplitude of the variable star. Each of the histograms is
consistent with an amplitude of 0.4 mag, as independently calculated
from the change in residuals around the PL relation, discussed in
Section 2.2.

\noindent

\begin{deluxetable}{lccccc}
\tablecolumns{6}
\tablewidth{4.5truein}
\tablehead{
\colhead{Cepheid}  & \colhead{~~~~~log(P)~~~~~}  & \colhead{3.6$\mu$m}    & \colhead{4.5$\mu$m}  & \colhead{5.8$\mu$m } & \colhead{8.0$\mu$m } 
\\ \colhead{}    & \colhead{(days)}    & \colhead{(mag)} & \colhead{(mag)}   & \colhead{(mag)} &\colhead{(mag)} \cr
}
\startdata
HV 872  & 1.475 & 11.174 & 11.245 & 11.167 & 11.129 \\
 &  & 0.034  & 0.033  & 0.041  & 0.053  \\
HV 872  & 1.475 & 11.397 & 11.481 & 11.373 & 11.350 \\
 &  & 0.023  & 0.026  & 0.032  & 0.046  \\
HV 873  & 1.536 & 10.764 & 10.891 & 10.740 & 10.701 \\
 &  & 0.040  & 0.035  & 0.034  & 0.052  \\
HV 873  & 1.536 & 10.733 & 10.773 & 10.721 & 10.658 \\
 &  & 0.034  & 0.029  & 0.040  & 0.057  \\
HV 875  & 1.482 & 11.062 & 11.011 & 11.030 & 11.027 \\
 &  & 0.034  & 0.025  & 0.034  & 0.034  \\
HV 875  & 1.482 & 11.020 & 10.971 & 10.946 & 10.939 \\
 &  & 0.039  & 0.031  & 0.039  & 0.038  \\
HV 876  & 1.356 & 11.579 & 11.687 & 11.624 & 11.527 \\
 &  & 0.037  & 0.032  & 0.055  & 0.063  \\
HV 876  & 1.356 & 11.537 & 11.538 & 11.516 & 11.575 \\
 &  & 0.036  & 0.038  & 0.040  & 0.076  \\
HV 877  & 1.655 & 10.584 & 10.687 & 10.606 & 10.571 \\
 &  & 0.035  & 0.036  & 0.039  & 0.050  \\
HV 877  & 1.655 & 10.595 & 10.687 & 10.634 & 10.542 \\
 &  & 0.040  & 0.027  & 0.033  & 0.041  \\
HV 878  & 1.367 & 11.387 & 11.387 & 11.277 & 11.318 \\
 &  & 0.035  & 0.041  & 0.042  & 0.041  \\
HV 878  & 1.367 & 11.373 & 11.471 & 11.375 & 11.402 \\
 &  & 0.041  & 0.037  & 0.041  & 0.057  \\
HV 879  & 1.566 & 10.718 & 10.801 & 10.725 & 10.708 \\
 &  & 0.024  & 0.024  & 0.031  & 0.035  \\
HV 879  & 1.566 & 10.904 & 10.821 & 10.824 & 10.811 \\
 &  & 0.037  & 0.023  & 0.041  & 0.032  \\
HV 882  & 1.503 & 11.087 & 11.139 & 11.014 & 10.969 \\
 &  & 0.032  & 0.033  & 0.039  & 0.046  \\
HV 882  & 1.503 & 11.315 & 11.338 & 11.215 & 11.163 \\
 &  & 0.053  & 0.034  & 0.036  & 0.050  \\
HV 887  & 1.161 & 12.288 & 12.217 & 12.213 & 12.223 \\
 &  & 0.052  & 0.074  & 0.065  & 0.056  \\
HV 887  & 1.161 & 12.235 & 12.238 & 12.233 & 12.197 \\
 &  & 0.039  & 0.042  & 0.050  & 0.057  \\
HV 889  & 1.412 & 11.189 & 11.342 & 11.228 & 11.119 \\
 &  & 0.042  & 0.060  & 0.050  & 0.053  \\
HV 889  & 1.412 & 11.226 & 11.267 & 11.253 & 11.170 \\
 &  & 0.058  & 0.043  & 0.048  & 0.053  \\
HV 891  & 1.235 & 11.734 & 11.721 & 11.788 & 11.652 \\
 &  & 0.041  & 0.042  & 0.049  & 0.059  \\
HV 891  & 1.235 & 12.036 & 12.082 & 12.055 & 12.034 \\
 &  & 0.051  & 0.047  & 0.070  & 0.058  \\
HV 892  & 1.204 & 12.155 & 12.148 & 12.096 & 12.194 \\
 &  & 0.048  & 0.038  & 0.042  & 0.066  \\
HV 892  & 1.204 & 11.982 & 11.959 & 11.930 & 11.967 \\
 &  & 0.038  & 0.033  & 0.051  & 0.056  \\
HV 893  & 1.325 & 11.730 & 11.705 & 11.674 & 11.706 \\
 &  & 0.051  & 0.037  & 0.065  & 0.053  \\
HV 893  & 1.325 & 11.697 & 11.776 & 11.684 & 11.675 \\
 &  & 0.044  & 0.042  & 0.056  & 0.051  \\
HV 899  & 1.492 & 11.350 & 11.464 & 11.359 & 11.275 \\
 &  & 0.032  & 0.039  & 0.049  & 0.049  \\
HV 899  & 1.492 & 11.213 & 11.219 & 11.153 & 11.137 \\
 &  & 0.035  & 0.042  & 0.068  & 0.041  \\
HV 900  & 1.677 & 10.324 & 10.350 & 10.311 & 10.186 \\
 &  & 0.027  & 0.029  & 0.034  & 0.044  \\
HV 900  & 1.677 & 10.382 & 10.441 & 10.353 & 10.263 \\
 &  & 0.023  & 0.029  & 0.028  & 0.043  \\
HV 901  & 1.266 & 11.870 & 11.919 & 11.910 & 11.773 \\
 &  & 0.039  & 0.040  & 0.044  & 0.071  \\
HV 901  & 1.266 & 12.213 & 12.202 & 12.198 & 12.037 \\
 &  & 0.028  & 0.026  & 0.041  & 0.041  \\
HV 904  & 1.483 & 11.386 & 11.331 & 11.318 & 11.271 \\
 &  & 0.025  & 0.021  & 0.036  & 0.035  \\
HV 904  & 1.483 & 11.171 & 10.976 & 11.001 & 10.937 \\
 &  & 0.045  & 0.038  & 0.042  & 0.034  \\
HV 909  & 1.575 & 10.951 & 10.981 & 10.943 & 10.922 \\
 &  & 0.036  & 0.040  & 0.039  & 0.047  \\
HV 909  & 1.575 & 10.661 & 10.686 & 10.586 & 10.580 \\
 &  & 0.042  & 0.041  & 0.040  & 0.039  \\
HV 914  & 0.838 & 13.046 & 13.052 & 13.145 & 12.927 \\
 &  & 0.034  & 0.051  & 0.05   & 0.097  \\
HV 914  & 0.838 & 13.226 & 13.204 & 13.195 & 13.320 \\
 &  & 0.042  & 0.036  & 0.095  & 0.109  \\
HV 971  & 0.968 & 12.808 & 12.806 & 12.808 & 12.795 \\
 &  & 0.042  & 0.053  & 0.082  & 0.088  \\
HV 971  & 0.968 & 12.638 & 12.504 & 12.588 & 12.552 \\
 &  & 0.048  & 0.038  & 0.061  & 0.080  \\
HV 932  & 1.123 & 12.213 & 12.167 & 12.165 & 11.935 \\
 &  & 0.039  & 0.050  & 0.068  & 0.145  \\
HV 932  & 1.123 & 12.415 & 12.458 & 12.313 & 12.057 \\
 &  & 0.031  & 0.040  & 0.051  & 0.103  \\
HV 953  & 1.680 & 10.159 & 10.219 & 10.167 & 10.135 \\
 &  & 0.030  & 0.032  & 0.043  & 0.041  \\
HV 953  & 1.680 & 10.266 & 10.364 & 10.234 & 10.185 \\
 &  & 0.037  & 0.029  & 0.033  & 0.043  \\
HV 997  & 1.119 & 12.403 & 12.550 & 12.484 & 12.376 \\
 &  & 0.043  & 0.041  & 0.071  & 0.065  \\
HV 997  & 1.119 & 12.233 & 12.229 & 12.177 & 12.175 \\
 &  & 0.032  & 0.038  & 0.053  & 0.059  \\
HV 1002 & 1.484 & 10.900 & 10.930 & 10.898 & 10.814 \\
 &  & 0.032  & 0.030  & 0.046  & 0.036  \\
HV 1002 & 1.484 & 11.204 & 11.257 & 11.141 & 11.120 \\
 &  & 0.037  & 0.041  & 0.037  & 0.046  \\
HV 1003 & 1.387 & 11.511 & 11.499 & 11.502 & 11.483 \\
 &  & 0.034  & 0.033  & 0.044  & 0.061  \\
HV 1003 & 1.387 & 11.505 & 11.453 & 11.530 & 11.469 \\
 &  & 0.043  & 0.033  & 0.042  & 0.051  \\
HV 1005 & 1.272 & 11.787 & 11.780 & 11.722 & 11.741 \\
 &  & 0.032  & 0.028  & 0.038  & 0.045  \\
HV 1005 & 1.272 & 12.120 & 12.162 & 12.035 & 12.137 \\
 &  & 0.035  & 0.049  & 0.057  & 0.073  \\
HV 1006 & 1.153 & 12.113 & 12.205 & 12.152 & 12.020 \\
 &  & 0.047  & 0.040  & 0.067  & 0.053  \\
HV 1006 & 1.153 & 12.114 & 12.214 & 12.056 & 12.054 \\
 &  & 0.052  & 0.059  & 0.066  & 0.058  \\
HV 1013 & 1.383 & 11.515 & 11.598 & 11.508 & 11.412 \\
 &  & 0.024  & 0.026  & 0.031  & 0.038  \\
HV 1013 & 1.383 & 11.611 & 11.606 & 11.590 & 11.506 \\
 &  & 0.041  & 0.033  & 0.051  & 0.057  \\
HV 1019 & 1.134 & 12.262 & 12.279 & 12.240 & 12.348 \\
 &  & 0.043  & 0.042  & 0.054  & 0.099  \\
HV 1019 & 1.134 & 12.316 & 12.287 & 12.310 & 12.365 \\
 &  & 0.038  & 0.031  & 0.052  & 0.064  \\
HV 1023 & 1.424 & 11.306 & 11.445 & 11.334 & 11.249 \\
 &  & 0.035  & 0.031  & 0.037  & 0.044  \\
HV 1023 & 1.424 & 11.174 & 11.275 & 11.196 & 11.150 \\
 &  & 0.027  & 0.036  & 0.041  & 0.050  \\
HV 2244 & 1.145 & 12.321 & 12.333 & 12.367 & 12.273 \\
 &  & 0.052  & 0.043  & 0.065  & 0.064  \\
HV 2244 & 1.145 & 12.241 & 12.174 & 12.173 & 12.165 \\
 &  & 0.045  & 0.031  & 0.058  & 0.062  \\
HV 2251 & 1.446 & 11.094 & 11.011 & 11.011 & 11.019 \\
 &  & 0.027  & 0.045  & 0.037  & 0.042  \\
HV 2251 & 1.446 & 11.110 & 11.176 & 11.151 & 11.088 \\
 &  & 0.031  & 0.036  & 0.032  & 0.030  \\
HV 2257 & 1.594 & 10.985 & 11.011 & 10.914 & 10.907 \\
 &  & 0.038  & 0.041  & 0.039  & 0.048  \\
HV 2257 & 1.594 & 10.521 & 10.590 & 10.526 & 10.526 \\
 &  & 0.044  & 0.031  & 0.036  & 0.046  \\
HV 2260 & 1.114 & 12.458 & 12.457 & 12.399 & 12.463 \\
 &  & 0.041  & 0.035  & 0.045  & 0.051  \\
HV 2260 & 1.114 & 12.797 & 12.808 & 12.870 & 12.684 \\
 &  & 0.065  & 0.039  & 0.080  & 0.090  \\
HV 2270 & 1.134 & 12.398 & 12.309 & 12.305 & 12.265 \\
 &  & 0.047  & 0.034  & 0.063  & 0.068  \\
HV 2270 & 1.134 & 12.318 & 12.308 & 12.293 & 12.174 \\
 &  & 0.054  & 0.029  & 0.055  & 0.058  \\
HV 2282 & 1.166 & 12.082 & 12.087 & 11.989 & 12.063 \\
 &  & 0.066  & 0.042  & 0.051  & 0.068  \\
HV 2282 & 1.166 & 12.128 & 12.125 & 12.120 & 12.144 \\
 &  & 0.051  & 0.041  & 0.059  & 0.069  \\
HV 2291 & 1.349 & 11.902 & 11.863 & 11.821 & 11.792 \\
 &  & 0.040  & 0.046  & 0.066  & 0.052  \\
HV 2291 & 1.349 & 11.478 & 11.499 & 11.462 & 11.434 \\
 &  & 0.035  & 0.041  & 0.045  & 0.037  \\
HV 2294 & 1.563 & 10.526 & 10.494 & 10.422 & 10.477 \\
 &  & 0.047  & 0.035  & 0.043  & 0.047  \\
HV 2294 & 1.563 & 10.808 & 10.765 & 10.726 & 10.759 \\
 &  & 0.035  & 0.029  & 0.044  & 0.036  \\
HV 2324 & 1.160 & 12.159 & 12.133 & 12.087 & 12.130 \\
 &  & 0.048  & 0.036  & 0.051  & 0.056  \\
HV 2324 & 1.160 & 12.325 & 12.276 & 12.215 & 12.313 \\
 &  & 0.046  & 0.040  & 0.056  & 0.077  \\
HV 2337 & 0.837 & 13.159 & 13.189 & 13.251 & 13.195 \\
 &  & 0.046  & 0.068  & 0.072  & 0.113  \\
HV 2337 & 0.837 & 13.230 & 13.194 & 13.070 & 13.321 \\
 &  & 0.051  & 0.070  & 0.088  & 0.129  \\
HV 2338 & 1.625 & 10.759 & 10.688 & 10.711 & 10.690 \\
 &  & 0.026  & 0.037  & 0.042  & 0.038  \\
HV 2338 & 1.625 & 10.370 & 10.445 & 10.389 & 10.347 \\
 &  & 0.028  & 0.020  & 0.031  & 0.031  \\
HV 2339 & 1.142 & 12.315 & 12.245 & 12.267 & 12.247 \\
 &  & 0.045  & 0.035  & 0.066  & 0.073  \\
HV 2339 & 1.142 & 12.101 & 12.084 & 12.078 & 12.113 \\
 &  & 0.029  & 0.025  & 0.042  & 0.054  \\
HV 2352 & 1.134 & 12.257 & 12.277 & 12.223 & 12.295 \\
 &  & 0.082  & 0.034  & 0.066  & 0.090  \\
HV 2352 & 1.134 & 12.393 & 12.341 & 12.298 & 12.337 \\
 &  & 0.041  & 0.027  & 0.065  & 0.062  \\
HV 2405 & 0.840 & 13.402 & 13.335 & 13.344 & 13.366 \\
 &  & 0.047  & 0.036  & 0.073  & 0.101  \\
HV 2405 & 0.840 & 13.250 & 13.296 & 13.109 & 13.171 \\
 &  & 0.039  & 0.041  & 0.079  & 0.097  \\
HV 2369 & 1.684 & 10.190 & 10.168 & 10.128 & 10.125 \\
 &  & 0.039  & 0.036  & 0.035  & 0.049  \\
HV 2369 & 1.684 & 10.133 & 10.169 & 10.109 & 10.083 \\
 &  & 0.038  & 0.029  & 0.040  & 0.044  \\
HV 2432 & 1.038 & 12.459 & 12.438 & 12.530 & 12.414 \\
 &  & 0.069  & 0.038  & 0.068  & 0.072  \\
HV 2432 & 1.038 & 12.488 & 12.352 & 12.389 & 12.482 \\
 &  & 0.045  & 0.043  & 0.059  & 0.083  \\
HV 2527 & 1.112 & 12.669 & 12.610 & 12.562 & 12.530 \\
 &  & 0.028  & 0.043  & 0.043  & 0.078  \\
HV 2527 & 1.112 & 12.461 & 12.519 & 12.447 & 12.402 \\
 &  & 0.036  & 0.033  & 0.071  & 0.066  \\
HV 2538 & 1.142 & 12.219 & 12.245 & 12.153 & 12.083 \\
 &  & 0.030  & 0.020  & 0.046  & 0.038  \\
HV 2538 & 1.142 & 12.204 & 12.318 & 12.298 & 12.201 \\
 &  & 0.034  & 0.028  & 0.033  & 0.053  \\
HV 2549 & 1.209 & 11.738 & 11.759 & 11.738 & 11.718 \\
 &  & 0.039  & 0.041  & 0.056  & 0.056  \\
HV 2549 & 1.209 & 11.797 & 11.819 & 11.757 & 11.757 \\
 &  & 0.030  & 0.028  & 0.052  & 0.046  \\
HV 2579 & 1.128 & 12.104 & 12.121 & 12.078 & 12.075 \\
 &  & 0.038  & 0.034  & 0.056  & 0.079  \\
HV 2579 & 1.128 & 12.263 & 12.159 & 12.138 & 12.193 \\
 &  & 0.051  & 0.031  & 0.058  & 0.051  \\
HV 2580 & 1.228 & 11.721 & 11.821 & 11.763 & 11.707 \\
 &  & 0.028  & 0.023  & 0.032  & 0.038  \\
HV 2580 & 1.228 & 11.717 & 11.727 & 11.711 & 11.618 \\
 &  & 0.038  & 0.034  & 0.054  & 0.048  \\
HV 2733 & 0.941 & 12.856 & 12.820 & 12.848 & 12.812 \\
 &  & 0.050  & 0.039  & 0.057  & 0.092  \\
HV 2733 & 0.941 & 12.814 & 12.828 & 12.823 & 12.900 \\
 &  & 0.036  & 0.033  & 0.058  & 0.065  \\
HV 2749 & 1.364 & 11.720 & 11.768 & 11.690 & 11.543 \\
 &  & 0.045  & 0.039  & 0.058  & 0.063  \\
HV 2749 & 1.364 & 11.763 & 11.645 & 11.616 & 11.598 \\
 &  & 0.032  & 0.038  & 0.043  & 0.070  \\
HV 2793 & 1.283 & 11.529 & 11.626 & 11.561 & 11.516 \\
 &  & 0.031  & 0.044  & 0.053  & 0.055  \\
HV 2793 & 1.283 & 11.699 & 11.784 & 11.621 & 11.644 \\
 &  & 0.047  & 0.036  & 0.048  & 0.062  \\
HV 2836 & 1.244 & 11.962 & 12.037 & 11.908 & 12.055 \\
 &  & 0.031  & 0.034  & 0.061  & 0.053  \\
HV 2836 & 1.244 & 11.953 & 12.109 & 11.973 & 11.943 \\
 &  & 0.037  & 0.041  & 0.057  & 0.064  \\
HV 2854 & 0.936 & 12.942 & 12.886 & 12.940 & 12.880 \\
 &  & 0.029  & 0.029  & 0.048  & 0.067  \\
HV 2854 & 0.936 & 12.779 & 12.743 & 12.684 & 12.798 \\
 &  & 0.034  & 0.028  & 0.051  & 0.072  \\
HV 5655 & 1.153 & 12.167 & 12.185 & 12.115 & 12.109 \\
 &  & 0.050  & 0.040  & 0.054  & 0.065  \\
HV 5655 & 1.153 & 12.120 & 12.150 & 12.149 & 12.110 \\
 &  & 0.038  & 0.030  & 0.044  & 0.035  \\
HV 6065 & 0.835 & 13.201 & 13.257 & 13.162 & 13.331 \\
 &  & 0.036  & 0.040  & 0.076  & 0.105  \\
HV 6065 & 0.835 & 13.414 & 13.365 & 13.310 & 13.285 \\
 &  & 0.033  & 0.045  & 0.078  & 0.101  \\
HV 6098 & 1.384 & 11.168 & 11.134 & 11.135 & 11.151 \\
 &  & 0.027  & 0.027  & 0.028  & 0.032  \\
HV 6098 & 1.384 & 11.189 & 11.126 & 11.086 & 11.089 \\
 &  & 0.035  & 0.037  & 0.044  & 0.046  \\
HV 8036 & 1.453 & 11.224 & 11.316 & 11.215 & 11.293 \\
 &  & 0.041  & 0.043  & 0.043  & 0.053  \\
HV 8036 & 1.453 & 11.360 & 11.466 & 11.385 & 11.287 \\
 &  & 0.042  & 0.025  & 0.055  & 0.038  \\
HV 12471& 1.200 & 12.044 & 12.129 & 12.030 & 12.007 \\
 &  & 0.040  & 0.035  & 0.053  & 0.067  \\
HV 12471& 1.200 & 12.253 & 12.339 & 12.256 & 12.223 \\
 &  & 0.034  & 0.041  & 0.060  & 0.057  \\
HV 12505& 1.158 & 12.416 & 12.453 & 12.533 & 12.460 \\
 &  & 0.036  & 0.035  & 0.085  & 0.074  \\
HV 12505& 1.158 & 12.539 & 12.499 & 12.407 & 12.646 \\
 &  & 0.044  & 0.036  & 0.061  & 0.085  \\
HV 12656& 1.127 & 12.339 & 12.299 & 12.289 & 12.301 \\
 &  & 0.048  & 0.037  & 0.058  & 0.067  \\
HV 12656& 1.127 & 12.217 & 12.187 & 12.145 & 12.198 \\
 &  & 0.056  & 0.041  & 0.054  & 0.078  \\
HV 12700& 0.911 & 12.910 & 12.908 & 12.989 & 12.899 \\
 &  & 0.043  & 0.044  & 0.073  & 0.085  \\
HV 12700& 0.911 & 12.900 & 12.914 & 12.873 & 12.996 \\
 &  & 0.035  & 0.030  & 0.066  & 0.071  \\
HV 12724& 1.138 & 12.326 & 12.365 & 12.338 & 12.316 \\
 &  & 0.051  & 0.039  & 0.064  & 0.071  \\
HV 12724& 1.138 & 12.492 & 12.566 & 12.463 & 12.443 \\
 &  & 0.046  & 0.040  & 0.073  & 0.063  \\
HV 12815& 1.416 & 11.063 & 11.125 & 11.059 & 11.007 \\
 &  & 0.038  & 0.034  & 0.042  & 0.038  \\
HV 12815& 1.416 & 11.054 & 11.079 & 11.004 & 11.053 \\
 &  & 0.030  & 0.024  & 0.029  & 0.033  \\
HV 12816& 0.973 & 12.897 & 12.844 & 12.901 & 12.956 \\
 &  & 0.052  & 0.034  & 0.075  & 0.077  \\
HV 12816& 0.973 & 12.826 & 12.789 & 12.877 & 12.765 \\
 &  & 0.035  & 0.036  & 0.061  & 0.085  \\
HV 13048& 0.836 & 13.084 & 13.088 & 13.067 & 13.054 \\
 &  & 0.034  & 0.032  & 0.059  & 0.075  \\
HV 13048& 0.836 & 13.217 & 13.197 & 13.060 & 13.058 \\
 &  & 0.044  & 0.044  & 0.090  & 0.093  \\
HV 2279 & 0.839 & 13.409 & 13.459 & 13.363 & 13.378 \\
 &  & 0.047  & 0.047  & 0.100  & 0.146  \\
HV 2279 & 0.839 & 13.346 & 13.472 & 13.463 & 13.186 \\
 &  & 0.067  & 0.059  & 0.105  & 0.109  \\
HV 886  & 1.380 & 11.488 & 11.550 & 11.481 & 11.450 \\
 &  & 0.034  & 0.033  & 0.048  & 0.055  \\
HV 886  & 1.380 & 11.667 & 11.232 & 11.552 & 11.585 \\
 &  & 0.039  & 0.035  & 0.052  & 0.055  \\
HV 12747& 0.556 & 14.205 & 14.251 & 14.527 & 14.051 \\
 &  & 0.044  & 0.058  & 0.174  & 0.185  \\
HV 12747& 0.556 & 14.224 & 14.193 & 14.135 & 13.999 \\
 &  & 0.034  & 0.041  & 0.091  & 0.999  \\
\enddata
\end{deluxetable}
\end{document}